\documentstyle[12pt]{article}
\def\beq{\begin{equation}}
\def\eeq{\end{equation}}

\begin{document}

\begin{titlepage}
\begin{center}
{\Large \bf Theoretical Physics Institute \\
University of Minnesota \\}  \end{center}
\vspace{0.3in}
\begin{flushright}
UMN-TH-1232/93 \\
TPI-MINN-93/61-T \\
December 1993
\end{flushright}
\vspace{0.4in}
\begin{center}
{\Large \bf $e^+\,e^- \to \tau^+ \, \tau^-$ at the threshold and beyond\\}
\vspace{0.2in}
{\bf Brian H. Smith \\}
School of Physics \& Astronomy, University of Minnesota \\
Minneapolis, MN 55455 \\
and \\
{\bf M.B. Voloshin  \\ }
Theoretical Physics Institute, University of Minnesota \\
Minneapolis, MN 55455 \\
and \\
Institute of Theoretical and Experimental Physics  \\
                         Moscow, 117259 \\
\vspace{0.2in}
{\bf   Abstract  \\ }
\end{center}

The excitation curve for the $\tau^+ \, \tau^-$ production in electron
positron annihilation near the threshold is revisited with the aim of
updating and extending a previous work. We find that the corrections of the
relative magnitude $O(\alpha)$ near the threshold are significantly
contributed by the radiative modification of the Coulomb interaction between
the $\tau$ leptons. The interpolation between the Coulomb effects at the
threshold and the relativistic effects at higher energies is considered and
the resulting formula is argued to have relative accuracy up to, but not
including, terms of the order of $\alpha^2$ at any velocity
of the $\tau$ leptons.
\end{titlepage}

The dramatic improvement in the accuracy of the measured mass of the $\tau$
lepton, achieved in the BES experiment$^{\cite{beps}}$, has demonstrated one
of the advantages of studying the production of the $\tau^+ \, \tau^-$ pair
in $e^+ \, e^-$ annihilation in the immediate vicinity of the threshold. In
view of the special kinematical and background advantages of the threshold
region$^{\cite{mv}}$ one may believe that experiments at the $\tau^+ \,
\tau^-$ threshold will be continued at the existing machine and, hopefully,
at the much anticipated tau-charm factory$^{\cite{pich}}$. In view of the
probable improvement in the precision of the experimental data, it is
worthwhile to have sufficiently accurate theoretical  description of the
$e^+\,e^- \to \tau^+ \, \tau^-$ cross section at the threshold and beyond.
In addition, this process is of a theoretical interest of its own, since it
provides a `test ground' for QED almost as good as the muon or electron
magnetic anomaly.

The well-known lowest-order QED formula for the cross section

\beq
\sigma_0 (e^+\,e^- \to \tau^+ \, \tau^-) ={{2 \pi \, \alpha^2} \over {3s}}\,
v\, (3-v^2)~~,
\label{s0}
\eeq
where $v=\sqrt{1-4 m_\tau^2/s}$ is the
velocity of each of the $\tau$ leptons in the c.m. frame, is subject to
corrections, arising from various sources: \\ {\it i} -- radiation from the
initial electron and positron, \\ {\it ii} -- vacuum polarization in the
time-like photon, \\ {\it iii} -- corrections to the spectral density $\rho
(q^2) = -{1 \over 3} \sum_X \langle 0 | j_\mu (-q) | X \rangle \langle X |
j_\mu (q) | 0 \rangle$ of the electromagnetic current $j_\mu = ({\bar \tau}
\, \gamma_\mu \tau)$ of the tau leptons. The latter corrections generally
include both the electromagnetic interaction between the $\tau$ leptons and
the final state radiation.

In higher orders there appears interference between the effects {\it i --
iii}, however its relative magnitude is suppressed by at least the factor
$\alpha^2$,which makes it beyond the scope of the present paper.

Because of the radiative effects the actually measured cross section is given
by the formula

\beq
\sigma(W)=\int^W \, r(W,w)\,|1-\Pi(w)|^{-2}\, {\bar \sigma}(w) \,dw~,
\label{se}
\eeq
where $W=\sqrt{s}$ is the total energy in the c.m. system. The weight
function $r(W,w)$ describes the radiation from the initial
state$^{\cite{kuraev}}$ and $|1-\Pi(w)|^{-2}$ is the vacuum polarization
factor$^{\cite{berends}}$. These two effects are standard and are routinely
accounted for in the data analyses, while the dynamics of the final state is
encoded in the cross section ${\bar \sigma}(w) = 8 \pi^2 \alpha^2 \rho (w^2)
/w^4$.

In this letter we concentrate on the behavior of ${\bar \sigma}(w)$ in the
threshold region, where the velocity $v$ of the produced $\tau$'s is a small
parameter. In this region the QED effects in the spectral density
$\rho(w^2)$ are determined by the Coulomb interaction between the $\tau$
leptons at distances $r \sim (m_\tau \,v)^{-1}$, the parameter for which
effects is given by $\pi \alpha /v$, while the relativistic effects and the
radiation of transversal photons by the $\tau$ leptons are suppressed by at
least the factor $v^2$. In the non-relativistic region one can effectively
separate the dynamics at short distances $r \sim m_\tau^{-1}$, at which the
leptons are produced by the electromagnetic current and that at the
distances $r \sim (m_\tau \,v)^{-1}$, which determine the Coulomb effects,
by replacing the current $j_\mu$ by a local non-relativistic operator. In
the lowest order in the loop corrections, arising at short distances, this
replacement can be written as (c.f. Ref.\cite{mv0})

\beq
{\bar \sigma}(w)= {{2 \pi^2 \alpha^2} \over {m_\tau^4}} {\rm Im}
G(0,0;m_\tau \, v^2)~~,
\label{nr0}
\eeq
where $G({\bf x},{\bf y};E)$ is the
non-relativistic Green's function at energy $E=w-2m_\tau$. For free
particles the Green's function is given by

\beq
G_0({\bf x},{\bf y};{{p^2} \over m}) = {m \over {4 \pi}} {{\exp (i\,p\,
|{\bf x}- {\bf y}|)} \over {|{\bf x}- {\bf y}|}}
\label{g0}
\eeq
so that ${\rm Im} G_0(0,0;m_\tau \, v^2)
= m^2 v /(4 \pi)$ and the non-relativistic limit of the `bare' cross section
(\ref{s0}) is reproduced. The solution for the Green's function in the
Coulomb potential $V(r)=-\alpha/r$ is also well known and at the origin it
reads as ${\rm Im} G_c(0,0;m_\tau \, v^2) = F_c\, {\rm Im} G_0(0,0;m_\tau \,
v^2)$ with the Coulomb factor $F_c$ given by

\beq
F_c={{\pi \alpha/v} \over {1-\exp (-\pi \alpha/v)}}~.
\label{fc}
\eeq
The $v^{-1}$ behavior of this factor at $v \to 0$ makes the cross section at
$v=0$ finite$^{\cite{mv}}$ and of a measurable magnitude:

\beq
{\bar \sigma} (v=0) ={{\pi^2 \alpha^3} \over {2 m_\tau^2}} \left ( 1 +
O(\alpha) \right ) = 0.236\,{\rm nb} \left ( 1 +
O(\alpha) \right ) ~.
\label{sv0}
\eeq

The fact that the spectral density of the current at small velocity is
determined by the Coulomb interaction calls for the idea that the radiative
corrections to this interaction should show up at the level of $O(\alpha)$
rather than the usual $O(\alpha^2)$. Since the higher order effects,
involving exchange and radiation of transversal photons at large distances,
are suppressed by the factor $v^2$, the only radiative effect at these
distances is due to running of the coupling $\alpha$, which is also known as
the Uehling-Serber radiative correction to the potential (see e.g. in the
textbook \cite{blp}). For a preliminary estimate of the magnitude of this
effect one can replace $\alpha$ in the Coulomb factor (\ref{fc}) by the
effective coupling at the momentum $p = m_\tau v$:

\beq
\alpha \to \alpha \, \left ( 1 + {{2 \alpha} \over {3 \pi}} \ln {{m_\tau v}
\over {m_e}} \right )~~.
\label{efc}
\eeq
One can readily see that the ratio $m_\tau v/m_e$ is large even for small
$v$, so that the effect can be of the order of one percent. This estimate
however requires some elaboration, which is the main subject of the present
work. Namely, non-logarithmic terms can be essential and also the estimate
of the characteristic distances, essential for the Coulomb factor, is
modified by the Coulomb interaction at $v \sim O(\alpha)$. To find the
effect quantitatively we use the full form of the Uehling-Serber correction
to the Coulomb potential due to the electron-positron vacuum
polarization$^{\cite{blp}}$:

\beq
\delta V(r) =  -{{2 \alpha^2} \over {3 \pi}}{1 \over r} \, \int_1^\infty \,
e^{-2 m_e r x} \left ( 1 + {1 \over {2x^2}}\right ){\sqrt{x^2-1} \over
{x^2}}\,dx~,
\label{usp}
\eeq
and consider this correction as a perturbation, thus finding the
modification of the wave function at the origin in the form

\beq
\delta \psi(0) = - \int G_c(0,r; m_\tau v^2)\, \delta V(r)\, \psi_c(r) \,d^3
r ~~,
\label{pt}
\eeq
where $\psi_c(r)$ is the S-wave wave function at energy $E=m_\tau v^2$ in
the Coulomb field $-\alpha/r$:

\beq
\psi_c(r)=C \, e^{-i p r} \, _1 F_1(1+ i \lambda, 2 , 2i p r)
\label{psic}
\eeq
with $p=m_\tau v$, $\lambda=m_\tau \alpha /(2p)= \alpha/(2v)$ and $C$ being
normalization constant. Using the representation of the Coulomb Green's
function in the form

\beq
G_c(0,r; p^2/m_\tau)=-i {{m \,p} \over {2 \pi}} \, e^{i p r}\,\int_0^\infty
e^{2 i p r t} \left ( {{1+t} \over t} \right )^{i \lambda} \, dt ~~,
\label{gfc}
\eeq
we find that the corrected expression for the square of the wave function at
the origin can be written as:

\beq
|\psi_c(0)+\delta \psi(0)|^2=|\psi_c(0)|^2 \, (1+h)~~,
\label{psicor}
\eeq
where the correction $h$ after integration over $r$ in eq.(\ref{psic}) is
given by

\beq
h= {{2 \alpha} \over {3 \pi}} \left [ - 2\lambda \, {\rm Im} \int_0^\infty
\,dt \, \int_1^\infty \, dx \left ( {{1+t} \over t} \right )^{i \lambda}
{{\left ( t+z \, x\,v^{-1} \right )^{i \lambda -1} }
\over
{\left ( t+1+z \, x\,v^{-1} \right )^{i \lambda +1} }}
\left ( 1 + {1 \over {2x^2}}\right ){\sqrt{x^2-1} \over
{x^2}} \right ]
\label{formula}
\eeq
with $z=m_e/m_\tau$.

The double integral in the latter equation was calculated
numerically{\footnote {In particular we found that the numerical integration
routines built into {\it Mathematica} provide a reliable, though not the
fastest, platform for this calculation.}}.
The plot of the magnitude of the relative correction $h$ as a function of
velocity $v$ is shown in Fig.1 along with the result of the simple
logarithmic estimate from the equations (\ref{fc}) and (\ref{efc}). One can
see that this estimate becomes invalid in the region $v < O(\alpha)$, while
it fits the exact result quite well at higher velocity. Also in Fig.1 is
shown the result of calculation of the contribution to $h$ of the muon
vacuum polarization, which, as expected, is quite small. We also find that
in the limit $v \to 0$ and $m_e/m_\tau \to 0$ the quantity in the square
brackets in eq.(\ref{formula}) is equal to 4, so that $h(v=0,m_e=0)= 8
\alpha /(3 \pi)$.

In order to have a full description of the $O(\alpha)$ correction to the
excitation of the $\tau^+\,\tau^-$ channel near the threshold one should
also take into account the finite renormalization of the electromagnetic
current, which comes from short distances $r \sim m^{-1}$, which is well
known$^{\cite{schwinger}}$ to result in the overall factor $(1-4 \alpha
/\pi)$, which, unlike the term $h$, does not depend on the velocity in the
threshold region, i.e. unless the terms $O(v^2)$ are taken into account.

Summarizing this discussion we find that the cross section for production of
non-relativistic $\tau$ leptons with velocity $v$ including the terms of the
relative magnitude $O(\alpha)$ is given by

\beq
{\bar \sigma}={{\pi^2 \alpha^3} \over {2 m_\tau^2}} {1 \over {1-\exp(-\pi \,
\alpha /v)}} \left (1- {{4 \alpha} \over \pi} + h \right )~~,
\label{thresh}
\eeq
where $h$ is determined by the running of the QED coupling $\alpha$ and the
dominant contribution to it due to the electron-positron vacuum polarization
is given by eq.(\ref{formula}) and which is plotted on Fig.1.

At the present accuracy of experimental measurement of $m_\tau$, which is
about 0.3 MeV$^{\cite{beps}}$, the $O(\alpha)$ corrections discussed here
can be completely ignored. A simple analysis shows that these terms become
essential for measurements with accuracy of the order of $10^{-2}$ MeV, at
which level one should also take into account the contribution of the
Coulomb bound states$^{\cite{mv}}$. For completeness we cite here that the
binding energy of the $n$-th level is

\beq
|E| = {{m_\tau \, \alpha^2} \over {4\, n^2}} \approx {{24\,KeV} \over {n^2}}
\eeq
and its $e^+e^-$ decay rate is given by

\beq \Gamma_{ee}={{m_\tau \, \alpha^5} \over {6 \, n^3}} \approx {{6.1\, eV}
\over {n^3}}~~.
\eeq

Equation (\ref{thresh}) is applicable in the non-relativistic limit, i.e.
the relative magnitude of the corrections to it is determined by $v^2$.
On the other hand the exact in $v$ formula of the first order in $\alpha$ is
known (see e.g. in Schwinger's textbook \cite{schwinger}):
${\bar \sigma} (e^+e^- \to \tau^+\tau^-) = \sigma_0 \, \left ( 1+
{\alpha \over \pi} S(v) \right )$, where $\sigma_0$ is the bare cross
section, eq.(\ref{s0}), and
\beq
\begin{array}{ll}
S(v)=&{1 \over v} \left \{ (1+v^2) \left [ {{\pi^2} \over 6} + \ln \left(
{{1+v} \over 2} \right ) \, \ln \left ( {{1+v} \over {1-v}} \right )+
2\, {\rm Li}_2 \left ( {{1-v} \over {1+v}} \right ) +
2\, {\rm Li}_2 \left( {{1+v} \over 2} \right ) - \right. \right.\\
& \left. 2\,
{\rm Li}_2 \left( {{1-v} \over 2} \right ) - 4 \, {\rm Li}_2(v) +
{\rm Li}_2(v^2) \right ] + \left [ {{11} \over 8} (1+v^2) - 3v+
{1 \over 2} {{v^4} \over {(3-v^2)}} \right ]\,
\ln \left ( {{1+v} \over {1-v}} \right )+ \\
& \left. 6v \, \ln \left( {{1+v} \over 2} \right ) - 4v \, \ln v +
{3 \over 4} v {{(5-3v^2)} \over {(3-v^2)}} \right \}
\end{array}
\label{schw}
\eeq
with ${\rm Li}_2(x) = -\int_0^x \ln (1-t) \, dt/t = \sum_{n=1}^\infty
x^n/n^2$.  The full relativistic formula of the first order in $\alpha$
develops at low velocity the inaccuracy of the order of $(\alpha/v)^2$,
which matches the inaccuracy $O(v^2)$ of the non-relativistic formula
(\ref{thresh}) at $v^2 \sim \alpha$, where the relative magnitude of the
error in either of these formulas is $O(v^2) \sim O(\alpha)$. One can
however write  an expression$^{\cite{mv}}$, which interpolates between the
non-relativistic formula (\ref{thresh}), summing all the terms of the form
$(\alpha/v)^k$ and $\alpha \, (\alpha/v)^k$, and the full formula of the
first order in $\alpha$. The resulting interpolating formula can be argued
to have relative accuracy $O(\alpha^2)$ at any velocity{\footnote {We take
this opportunity to correct some misprints and other minor flaws of the
discussion of the interpolating formula in Ref.\cite{mv}. These flaws
however do not affect the results of the data analysis in the BES
experiment$^{\cite{beps}}$, which used the formulas from Ref.\cite{mv}.}}

The interpolating formula has the form

\beq
{\bar \sigma} (e^+e^- \to \tau^+\tau^-) = \sigma_0\, (1+ h) \, F_c \,
\left (1+{\alpha \over \pi} S(v) - {{\pi \, \alpha} \over {2v}} \right )~~
\label{interp}
\eeq
where the lowest-order cross section $\sigma_0$ is given by eq.(\ref{s0}),
the Coulomb factor $F_c$ is in the equation (\ref{fc}), the correction term
$h$, given by eq.(\ref{formula}), accounts for running of $\alpha$ in the
Coulomb terms and the function $S(v)$ is written in eq.(\ref{schw}).
One can notice that in eq.(\ref{interp}) the first Coulomb term $\pi \alpha
/(2v)$ is subtracted from the full relativistic expression of the first
order in $\alpha$, since this term is contained in the factor $F_c$.
The factor $(1-4 \alpha/\pi)$, coming from finite renormalization of the
current at the threshold, which is present in the equation (\ref{thresh}),
in the interpolating formula is contained in the factor with $S(v)$.
Equation (\ref{interp}) sums all terms of the form $(\alpha/v)^k$ and
$\alpha \, (\alpha/v)^k$ as well as all terms linear in $\alpha$ with
arbitrary dependence on $v$.  Therefore the corrections to the
interpolating formula start with terms of the form
$\alpha^2 \, (\alpha/v)^k$ with $k \ge 0$, which are at least as small as
$O(\alpha^2)$ in comparison with those included in eq.(\ref{interp}),
independently of the velocity $v$.  Naturally, to also take into account the
terms of the relative order $\alpha^2$ one needs a full calculation of the
two-loop correction to the spectral density of the electromagnetic current
at arbitrary $v$ as well as a calculation of the interference between
the mechanisms {\it i -- iii} mentioned in the beginning of this paper.

This work was supported, in part, by the DOE grant DOE-AC02-83ER40105.

\newpage

{\large \bf Figure caption}\\[0.15in]
{\bf Figure 1.} The relative correction $h$ (eq.(\ref{formula})) to the
cross section of $\tau^+ \tau^-$ production near the threshold, arising from
modification of the Coulomb potential by the polarization of the
electron-positron vacuum (solid) and of the muon one (dashed) vs. the
velocity of the $\tau$ leptons in the c.m. system. Also is shown (dotted)
the logarithmic approximation for the electron-positron contribution to $h$
according to eq.(\ref{efc}).

 \end{document}